\documentclass[sigconf]{acmart}
\usepackage{hyperref}
\usepackage{graphicx} 
\usepackage{array} 

\newcommand{\code}[1]{\texttt{#1}}

\copyrightyear{2026}
\acmYear{2026}
\setcopyright{cc}
\setcctype{by}
\acmConference[SCA/HPCAsiaWS 2026]{SCA/HPCAsia 2026 Workshops: Supercomputing Asia and International Conference on High Performance Computing in Asia Pacific Region Workshops}{January 26--29, 2026}{Osaka, Japan}
\acmBooktitle{SCA/HPCAsia 2026 Workshops: Supercomputing Asia and International Conference on High Performance Computing in Asia Pacific Region Workshops (SCA/HPCAsiaWS 2026), January 26--29, 2026, Osaka, Japan}
\acmPrice{}
\acmDOI{10.1145/3784828.3786261}
\acmISBN{979-8-4007-2328-5/2026/01}

\begin{document}

\title{The LCLStream Ecosystem for Multi-Institutional Dataset Exploration}

\author{David Rogers}
\email{rogersdm@ornl.gov}
\affiliation{%
  \institution{NCCS, Oak Ridge Leadership Computing Facility}
  \city{Oak Ridge}
  \state{Tennessee}
  \country{USA}
}

\author{Valerio Mariani}
\email{valmar@slac.stanford.edu}
\affiliation{%
  \institution{LCLS, SLAC National Accelerator Laboratory}
  \city{Menlo Park}
  \state{California}
  \country{USA}
}
\author{Cong Wang}
\email{cwang31@slac.stanford.edu}
\affiliation{%
  \institution{LCLS, SLAC National Accelerator Laboratory}
  \city{Menlo Park}
  \state{California}
  \country{USA}
}
\author{Ryan Coffee}
\email{coffee@slac.stanford.edu}
\affiliation{%
  \institution{LCLS, SLAC National Accelerator Laboratory}
  \city{Menlo Park}
  \state{California}
  \country{USA}
}
\author{Wilko Kroeger}
\email{wilko@slac.stanford.edu}
\affiliation{%
  \institution{LCLS, SLAC National Accelerator Laboratory}
  \city{Menlo Park}
  \state{California}
  \country{USA}
}
\author{Murali Shankar}
\email{mshankar@slac.stanford.edu}
\affiliation{%
  \institution{LCLS, SLAC National Accelerator Laboratory}
  \city{Menlo Park}
  \state{California}
  \country{USA}
}
\author{Hans Thorsten Schwander}
\email{thorsten@slac.Stanford.edu}
\affiliation{%
  \institution{LCLS, SLAC National Accelerator Laboratory}
  \city{Menlo Park}
  \state{California}
  \country{USA}
}
\author{Tom Beck}
\email{becktl@ornl.gov}
\affiliation{%
  \institution{NCCS, Oak Ridge Leadership Computing Facility, supported by the US DOE Office of Science under Contract No. DE-AC05-00OR22725.}
  \city{Oak Ridge}
  \state{Tennessee}
  \country{USA}
}
\author{Frédéric Poitevin}
\email{fpoitevi@slac.stanford.edu}
\affiliation{%
  \institution{LCLS, SLAC National Accelerator Laboratory}
  \city{Menlo Park}
  \state{California}
  \country{USA}
}
\author{Jana Thayer}
\email{jana@slac.stanford.edu}
\affiliation{%
  \institution{LCLS, SLAC National Accelerator Laboratory, supported by the US DOE Office of Basic Energy Sciences under Contract No. DE-AC02-76SF00515.}
  \city{Menlo Park}
  \state{California}
  \country{USA}
}


\renewcommand{\shortauthors}{D. M. Rogers, V. Mariani, C. Wang et al.}

\begin{abstract}
We describe a new end-to-end experimental data streaming framework designed from the ground up to support new types of applications --
AI training, extremely high-rate X-ray time-of-flight analysis,
crystal structure determination with distributed processing,
and custom data science applications and visualizers yet to be created.
Throughout, we use design choices merging cloud microservices with traditional HPC batch execution models for security and flexibility.
This project makes a unique contribution to the DOE Integrated Research Infrastructure (IRI) landscape.  By creating a flexible, API-driven
data request service, we address a significant
need for high-speed data streaming sources for the X-ray science data analysis community.
With the combination of data request API, mutual authentication web security framework, job queue system, high-rate data buffer, and complementary nature to facility infrastructure, the LCLStreamer framework has prototyped and implemented several new paradigms critical for future generation experiments.
\end{abstract}






\maketitle


%


\section{ Introduction}

Autonomous experiment steering is already needed now to quickly adapt to
changing experimental conditions that guide the instrument to optimal operating regimes.
For example, machine-driven, Bayesian optimization of mechanical alignment settings
for the incoming X-ray waveguides achieves 5x faster time-to-calibration than manual,
human-driven optimization.\cite{roussel_bayesian_2024}

However, coupling High-Performance Computing (HPC) to external, on-line
data sources requires the convergence of several new capabilities:
data ontologies\cite{rajamohan_materials_2025} for the naming, processing, and storage of results at each level
(e.g. raw events, X-ray crystal images, detected scattering peak positions, or
electron detection times), experimental analysis frameworks for
interactively guiding (CPU/GPU-intensive) data exploration,
cross-facility coordination on co-scheduling of experiments and analysis,
web-accessible applications programming interfaces (API-s) for data producers
and HPC jobs, robust software for high-bandwidth data transmission,
and a programming language coordinating interlinked, multi-participant activities.

This work discusses the design of a set of integrated services for experimental data collection
and analysis that we developed. Together, they accomplish
several objectives that would be impossible with a single, monolithic program:
automated data collection, local fast data reduction, data streaming to academic or HPC facilities,
user-defined analysis routines, and long-term data storage for data re-use (e.g. replicating
studies or AI model training).

\subsection{ Data Streaming Framework}

Experimental data at the Linac Coherent Light Sources (LCLS)
is accessed using the psana1 and psana2 libraries
(for LCLS and LCLS-II respectively).\cite{damiani_linac_2016,thayer_massive_2024}
These libraries read and assemble events from a
running or archived experiment.
These libraries are flexible, performant, and scalable across nodes
using MPI, but must be used on site for administrative
and technical reasons.
The services that provide data access run user code and access databases
(containing information needed to pre-process events).  These actions can only be
performed by authenticated users from within the LCLS facility.

The streaming framework described in this work opens up these data
by providing local and remote distributed access to the psana1 and psana2 data outputs.
Preliminary timing results show data arrival at an HPC job running on
Oak Ridge's National Center for Computational Sciences just seconds after
collection at SLAC's detectors in Menlo Park.
The net result is increased data availability for the growing
community of experts whose work focuses on experimental steering, time-sensitive processing, and online analysis.\cite{rakitin_introduction_2020,brim_microservices_2024,mariani_it_2016}


\begin{figure*}
    \centering
    \includegraphics[width=3.5in]{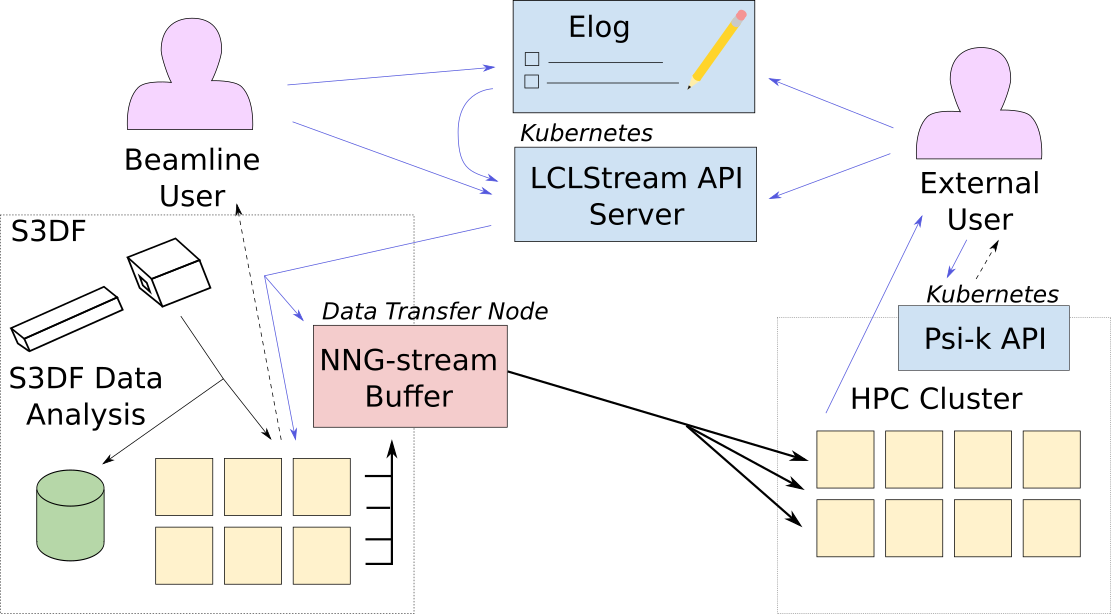}
    \caption{Data streaming process diagram.  Blue arrows show control paths, and black arrows show data flow.  Dotted paths are for returned results.  Event assembly and data formatting is performed by the psana framework on a compute job inside S3DF (left).  The LCLStream API can start network buffers (on a S3DF data transfer node) and compute jobs on S3DF to format and send experimental data. External users can pair an LCLStream API call with jobs on other HPC clusters (right).}
    \label{fig:core}
\end{figure*}

Fig.~\ref{fig:core} shows the core components of our data streaming framework.
The LCLStream API Server (in blue at the center of figure~\ref{fig:core}) allows
beamline or external users to request a dataset from a specific experiment
using a REST API with Javascript object notation
(JSON)-formatted queries.
The user performing the request and the server recognize each other
through a certificate-driven mutual authentication process.
LCLStream-API creates a JobID and launches a pair of processes
on S3DF (the SLAC Shared Science Data Facility).
First, it starts an NNG-Stream process a data transfer node
with a new pair of recieve/send ports for this stream.
Second, it starts LCLStreamer as an MPI job over SLURM on the
S3DF cluster (on the left in the same figure).
The receive URI is returned to the client.
All compute processes can make independent connections
to that address.

Together, the flexible data request API and the highly
configurable LCLStreamer application give users flexibility in selecting datasets, performing partial
data reduction, and choosing their preferred protocols and encodings.
To support the data transfer, we have also created NNG-Stream (red box).
It buffers data between parallel producers and consumers,
smoothing the data flow in case of bursts of network capacity or activity.
The buffer is stackable (not shown), so it can traverse complex network topologies.
Data can be sent to multiple types of external applications --
supercomputer centers, network-based appliances, and monitoring and automated control systems
(in yellow on the right).

In the sections that follow, we give details on several experiments that were run
using the early iterations of this infrastructure -- including electron time-of-flight
correlation analysis, image AI/ML training, and crystal structure inference from X-ray
scattering.  Then we go into greater detail on the interfaces, cache design,
integration with job scheduling, and key authentication and performance characteristics.
Overall, rapid progress has been made by keeping these individual components simple
-- speeding up the development cycle and decreasing the barrier for community
involvement.


\section{ Scientific Applications}

Application scientists are most interested in their specific problem domain.
However, almost every beamline has a different way to access its output,
along with metadata like detector positions, timestamps, and other instrument settings.
Users often have little inclination to learn a new data format
or data processing library.  For these reasons, users almost always start their work
by converting from the original data into their own formats.
Although the original structure and ways of working with the data are lost,
users are now in possession of an initial reduced dataset that is
much more useful for their work.

Given this historical experience, the OM (OnDA Monitor)
package for crystallographic data analysis has built a
specification for a data pre-processing
pipeline.\cite{mariani_it_2016}
In order to adapt this idea to LCLStreamer,
we considered the needs of several new and existing data analysis projects:
\begin{itemize}
  \item MAXIE - a masked autoencoder for X-ray Image reconstruction,
  \item PeakNet - an AI/ML method for peak-detection in crystal scattering X-ray images,
  \item TMO-prefex - a data reduction and fast histogram/correlation counting package used during the commissioning of the LCLS-II upgrade for the Time-resolved atomic, Molecular and Optical Science (TMO) beamline.
  \item CrystFEL - a program package for processing data collected in Serial Femtosecond Crystallography (SFX) experiments
\end{itemize}

Specifically, we looked for patterns in the data reduction steps and
the input data formats used by each application.
We also considered the needs of several developing projects which could make use of this data processing pipeline, including: online calibration estimation methods, high-rate
image transmission from the X-Ray pump-probe (XPP) hutch to GPU nodes for image processing,
and custom setups for CCTBX users who want more control over their
data formatting (e.g. transformation to NeXus data format).

Each of these projects has unique output types: foundational AI model
weights, reduced peak positions, histograms and experimental configuration-sensitive
summary statistics, and electron density maps.  However, all of them still bring
event output and metadata through a data processing pipeline with well-defined
processing steps (predictable compute requirements and intermediate result types).

At the beginning of our work, all of the above applications were able to run locally
at the S3DF, by reading xtc or xtc2-formatted event data
using LCLS's psana package.\cite{damiani_linac_2016,thayer_massive_2024}
These event data files are output by each experimental run,
can be generated at rates up to gigabytes per second (which will become terabytes in the future),
and are effectively stored/loaded/streamed in parallel by
LCLS's high-throughput hardware setup. 
As the project progressed, we merged the common steps in these applications,
and abstracted the differences into configurable options.
Most variations can now be handled by adding new input detectors
and data reduction functions, rather than rebuilding the entire data processing
application.

\subsection{ Masked X-ray Image Autoencoder (MAXIE) and PeakNet}

 Cong and Chen describe two AI models capable of reconstructing and interpreting
X-ray images.\cite{wang_end--end_2025,chen_augmenting_2025}
The team has developed the Masked Autoencoder for X-ray Image Encoding (MAXIE), supporting model architectures ranging from hundreds of millions to billions of parameters. MAXIE is trained on approximately 286 terabytes of X-ray diffraction images from a variety of different detectors.  Training epochs require hours to days depending on available compute resources, motivating its checkpointing and fault tolerance features.  The implementation supports multiple parallelization strategies within a unified training framework, including single GPU, multi-GPU, and multi-node configurations using both Distributed Data Parallel (DDP) and Fully Sharded Data Parallel (FSDP) approaches (including sharded and full checkpoints), with optimizations including shared memory utilization and job scheduler integration for fault-tolerant execution.

The goal for this application is to run unsupervised training of this model on the full 286-terabyte dataset,
potentially unlocking representations learned from the complete diversity of crystallographic data.
Current production results come from a complementary supervised learning approach operating on a significantly smaller, algorithmically labeled dataset.  Because both approaches leverage the same unified training infrastructure, scaling up to larger datasets is a matter of compute availability and network data accessibility. Architectural consistency in the AI/ML framework provides operational benefits including shared fault-tolerance mechanisms, unified checkpoint management, and the ability to scale supervised training to larger datasets as they become available.

\subsection{ TMO-prefex}

The LCLS-II upgrade for the Time-resolved atomic, Molecular and Optical Science (TMO) beamline
is designed to operate with X-ray laser shot repetition rates up to 1 MHz, collecting electron time of flight data with sub-femtosecond resolution after each shot.\cite{walter_multi-resolution_2021}

In theory, the increase in shot repetition rate from the previous 120 Hz allows
experiments to complete 8333 times faster.
However, the current state of the art for data analysis is running offline,
out of the main experimental path.  When done this way, it is not possible to
determine whether each experimental step has been correctly configured and informative
data has been gathered fast enough to keep up with the speed of data generation.
In order to actually accelerate the experimental
data collection phase, we need a fast data analysis pipeline.
Speeding up the iteration time needed for completing each experimental step is key to realizing 1000x productivity increases.
It will ultimately require data analysis
capable of reconfiguring the experimental parameters in an automated way.\cite{roussel_bayesian_2024}

One central output of the TMO beamline is electron time of flight detection
(FEX detector).  Electrons emitted by molecules trapped in the TMO
chamber are emitted at specific times after an initial laser excitation.
Because one molecule can emit several electrons, the times and angles
form a correlated signal, reporting on the molecule's relaxation process.

\begin{figure}
\includegraphics[width=\linewidth]{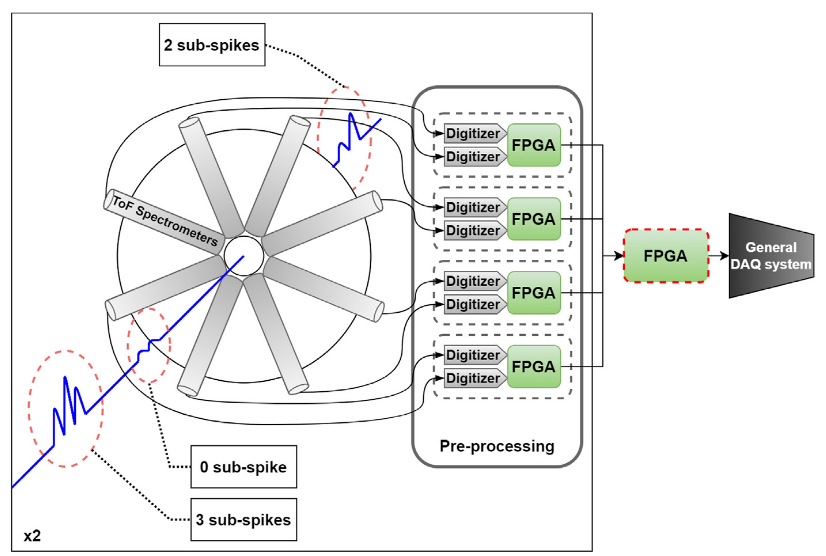}
\caption{TMO time of flight (ToF) detector configuration for detecting time and angular distribution of emitted electrons (reproduced from Ref.~\cite{gouin-ferland_data_2022}). 
ToF spectrometer signals are processed by analog electronics before being digitized. After event detection, the central FPGA (circled with a red dotted line), forwards event features from all 8 peripheral FPGA-s on to the S3DF data processing pipeline.
Instead of a custom code juggling event extraction with
LCLS-II's internal data path,
LCLStream simplifies data extraction from this source
down to a list of detector names and compression options.
}\label{fig:cookiebox}
\end{figure}

Figure~\ref{fig:cookiebox} shows the dataflow for signal acquisition and initial reduction from the electron time-of-flight detectors.\cite{gouin-ferland_data_2022}
Downstream processing is then used to take these raw arrival times and spectra and accumulate histograms of the electron arrival times and light spectra, as well as perform more detailed sorting and chain-of-event reconstruction.  These output histograms are the basis for Angle-resolved photo- and Auger–Meitner electron spectroscopy analyses (ARPES and ARAES, respectively).

The data stream output by the detectors is a current read-out for all
angular channels at all times with femtosecond resolution.
This extremely large data stream is eventually compressed into a
list of individual electron arrival times.  Three important intermediate steps
are 1) raw waveform data, 2) a compressed set of waveforms above a threshold value,
and 3) the final arrival times and detector numbers of current peaks.

Analysis on each type of data has historically been accomplished using ipython notebooks
that parse data from hdf5-formatted dictionaries of arrays (several named arrays per detector).
Even with good data tracking methods, it quickly becomes difficult to manage the combination of detector configurations, configuration options for processing steps, and conclusions made from analysis output.  More importantly, this process has a built-in latency because it cannot be automated.

We incorporated the full data collection, transformation, and output to HDF5
pipeline into LCLStreamer by modeling the FEX detector, the three types
of compression as data processing steps,
and the output to HDF5 using an HDF5 data serializer.
In our tests and evaluation of this framework, we
simultaneously started 128 MPI-parallel data producers across 2 nodes of the S3DF data analysis cluster, an instance of NNG-Stream on a data transfer node of S3DF, and a 8 MPI-parallel data receivers across two nodes of OLCF's ACE testbed.

The combination of steps above demonstrated important advantages in automation, metadata tracking, and workflow re-usability.
Experimental configurations are documented in the run log curated by SLAC's Elog system, as usual.
Data processing pipeline is captured both by LCLStreamer's configuration file and, when source
changes are needed, by its git version control.
Parallel processing, setup and tuned by facility staff, is used both for data output at S3DF and analysis at the OLCF HPC site.  Finally, API-driven
workflows substantially increase the accessibility and transparency of the above steps.

\subsection{ CrystFEL}

CrystFEL \cite{white_crystfel_2012} is a collection of programs for processing data collected using the Serial Femtosecond Crystallography technique. It includes programs performing all the steps needed to turn a set of diffraction images in a list of structure factors. This includes utilities to index and integrate diffraction patterns, and to merge measured intensities.

Live data analysis as the user is taking data at a beamline
is a killer feature because it allows the user to maximize
the scientific output from their beamtime.\cite{cctbx}
For example, determining whether detectors are aligned,
diffraction spot intensities cover a large enough
dynamic range, or whether the crystal itself contains growth
defects strongly impact the timeline of an experiment.

In support of this, CrystFEL can operate both on images stored in HDF5 format\cite{hdf5} in local files, or streamed via a network socket in the same format. Typically a data packet includes a diffraction image, and supplemental information needed for its interpretation (an estimation of the distance between the detector and the interaction point of the experiment, approximate beam energy, optical laser status, etc). Some additional pertinent experimental information (detector geometry, space group symmetry, etc.) is instead provided via program configuration files or command-line parameters.
A graphical front-end to control the processing is available.

Previous works showing live, HPC-assisted data analysis
mirrored raw (xtc) data files to the remote facility and then
ran the full psana-based software load process.\cite{cctbx24}
In the present work, we aim to do the extraction and compression
in parallel at the source.
Separating the data pipeline this way is somewhat simplified
by the program's ability to read source data from the network.
However, it is still challenging to arrange sending of the extra
metadata while starting and stopping the program on an HPC center.
If successful, the receiving
analysis software will be less complicated and the entire pipeline
will run with lower latency and higher throughput.

%
%

\section{ Software Infrastructure}

In order to support the types of applications described above,
we have constructed the LCLStream data pipeline from a modular
set of network services.
\begin{itemize}
\item \href{https://slac-lcls.github.io/lclstreamer}{LCLStreamer}: An engine for fast, flexible data reduction and formatting
\item LCLStream-API: An HTTPS-REST API wrapping LCLStreamer and NNG-Stream
\item \href{https://gitlab.com/frobnitzem/nng_stream}{NNG-Stream}: A buffer for receive-once, send-once message distribution from a concurrent set of producers to a concurrent set of consumers 
\item \href{https://pswww.slac.stanford.edu/}{Elog}: LCLS's electronic logbook, which enables run tracking and automated tasks
\item \href{https://github.com/frobnitzem/psik}{Psi-k} and \href{https://frobnitzem.github.io/psik_api/}{Psik-API}: A portable batch submission interface for jobs with HTTPS-REST API job and file access
\item \href{https://certified.readthedocs.io/}{certified}: An x.509 public key infrastructure for secure, mutually authenticated HTTPS client/server communication
\end{itemize}

LCLStreamer was designed based on patterns proven in the OM X-ray analysis software suite.\cite{mariani_it_2016}
This package has been developed over several years and has a strong user base within the
X-ray science community.  Its interface is designed around a top-level specification
file for the data collection, reduction, and analysis pipeline.  Users are able to achieve
flexibility by plugging in different producers, data processing steps,
and consumers within that pipeline.

The LCLStream API is a simple API wrapper starting and stopping LCLStream data producers.
This enables off-site users to send an API request and then begin receiving data.

NNG-Stream is a message buffer using the nanomsg next generation library.
It is designed to be simple and high-performance, operating by storing messages
in memory and sending them in first-in-first-out order.  It does no inspection
of message contents.

Elog has been LCLS's working database for many years.  It tracks experiment metadata
like run parameters, start and stop times, comments from beamline users, and other diagnostic
and analysis steps performed as part of a given experiment.  The usual workflow for experimental investigations involves setting up instrument parameters, starting and stopping data collection (creating a "run"), and performing post-experiment annotation and analysis to the resulting run entry in the Elog.  This operation makes Elog a natural place to store run-associated data, as it becomes a central location for organizing and automating runs forming an experiment and their associated run events.


Psi-k is a front-end to a file folder structure storing one job per folder.  It creates new job folders from a JobSpec data structure (parsed from json or yaml).  Job scripts can be created for different backends, including SLURM.  Each job script runs \code{psik reached} to record its progress through a state sequence (\code{queued} $\to$ \code{active} $\to$ \code{completed/canceled/failed}).

The state sequence and the format of the data structures used by Psi-k benefit strongly from specifications developed by the ExaWorks PSI/J project.\cite{psij}
Psi-k's design diverged, however, in order to provide web interoperability.
For example, Psi-k does not run job status checks by polling SLURM commands.
State changes are stored in a status file, and can also trigger webhooks.
Stdout and stderr are captured in a logs folder, and numbered sequentially for
each re-run of the job. These file layouts form the basis for exposing jobs via an API (Psik-API).  Mapping job create-read-update-destroy operations to the file hierarchy provides a straightforward REST-HTTPS interface to Psi-k.



\subsection{ Fast Data Reduction with LCLStreamer}

LCLStreamer exports LCLS's psana-native events (xtc/xtc2 format) to custom formats
as needed by SLAC beamline users.  It does this using the full parallelism supplied
by LCLS's data processing pipeline, applying all LCLS-related corrections and calibration.
Key for our users, it has a flexible processing and formatting step. It outputs data
suitable for direct consumption by the user's preferred external software.
Data from several events is accumulated, serialized, and finally passed to
handlers to save as a file, network stream (or both) to external applications.

The LCLStreamer application is made up of several parts: 

\begin{itemize}
\item An Event Source (that generates the data)
\item A Processing Pipeline (that performs data reduction, calibration, etc.)
\item A Serializer (that turns the data into a binary object)
\item One or more Data Handlers (that write the binary object to a file, sends it through a network sockets, etc)
\end{itemize}


The LCLStreamer pipeline starts by retrieving a single event from an \code{EventSource},
and extracts from the event all data requested by the user.
Any remaining data in the event is discarded.
The data retrieved for each event has the format of a Python dictionary
of Numpy Arrays. Each key in the dictionary corresponds to a data source.
The value associated with the key is the information retrieved
from the data source for the event being processed.

The operations of a \code{ProcessingPipeline} are then applied to
the data retrieved from each event.  This step makes use of the
\href{https://github.com/frobnitzem/stream.py}{stream.py}
library to compose together a series of python generators (coroutines).

The standard pipeline batches together the results
of processing several consecutive events.
This accomplishes the same kind of batching one sees in
a pytorch \code{DataLoader}, a key enabler of performance
and interoperability.
At that point, the accumulated data is returned in bulk.
The data still has the format of a dictionary of Numpy array,
with each key representing a data entry, and the corresponding value storing the accumulated data. The data is then serialized into a binary form.

Finally, the data is passed to one or more \code{DataHandler}s that can forward
the data to the filesystem or any other external application,
often via a network on in-process kernel socket, optionally in a compressed form.
If multiple \code{DataHandler}s are present, they handle the same binary blob in parallel.
New \code{Serializers} and \code{DataHandlers} can be added so that external applications
can consume the data format they prefer. 

Since there are multiple choices
for event sources, processing pipelines, data serializers, etc.,
each section of the configuration file is marked with a
\textit{type} to select which implementation class it will use.


Next to the type selection, additional configuration options can also
be provided to each of these implementation classes.
For example, the \code{HDF5Serializer} class implements
the data serializer interface.  It serializes its input data
into a binary string with the internal structure of an HDF5 file.
Its configuration options can defined be when the
\textit{data\_serializer} section selects the
\textit{HDF5Serializer} type.


\begin{verbatim}
data_serializer:
    type: HDF5Serializer
    compression_level: 3
    compression: zfp
    fields:
        timestamp: /data/timestamp
        detector_data: /data/data
\end{verbatim}

Finally, the \textit{data\_sources} section of the configuration file defines the data that LCLStreamer extracts from every data event it processes. If a piece of information is part of the data event, but not included in the \textit{data\_sources} section, LCLStreamer will ignore it (filtering at read time).
The \textit{data\_sources} section of the configuration file consists of a dictionary of data sources.
Each entry's key acts as a variable name that identifies
the extracted data throughout the whole LCLStreamer data workflow.
The entry's associated dictionary defines the nature of the
data source (via the mandatory \textit{type} entry) and any
other parameters needed to configure the initial data extraction from
the psana library.
The type of a data source is the name of a
Python class within LCLStreamer that implements extraction.

For example, the following blocks declares variables
\textit{timestamp} and \textit{detector\_data}.

\begin{verbatim}
data_sources:
    timestamp:
        type: Psana1Timestamp

    detector_data:
        type: Psana1AreaDetector
        psana_name: Jungfrau1M
        calibration: true
\end{verbatim}

The \textit{timestamp} data class is of type \code{Psana1Timestamp}.
Inside LCLStreamer, the \code{Psana1Timestamp} class will be called to
read the associated timestamp data for each event.
The \textit{detector\_data} class is instead of type \code{Psana1AreaDetector}.
The two configuration parameters \textit{psana\_name} and \textit{calibration}
are passed to the Python class \code{Psana1AreaDetector} that defines how this type of data is retrieved.

LCLStreamer currently supports psana \cite{damiani_linac_2016} and
psana2 \cite{thayer_massive_2024} as data sources, with a wide array of detectors and instrument
readouts.  It allows data to be serialized in customizable formats like HDF5 and can write
them to files and/or send them via \href{https://zeromq.org/}{ZMQ} or \href{https://nng.nanomsg.org/}{NNG} sockets.

\subsection{ Data availability with LCLStream-API}

LCLStream-API provides a web interface for external users to
request data from LCLStreamer.
It is built around LCLStreamer's configuration file.
As a REST-API, data transfers are started by POST operation,
sending the configuration
file as a typed JSON message to the \code{transfers} path.
The response is either a validation error, or the ID for the newly created
transfer.  Issuing a GET or a DELETE to \code{transfers/ID} then reads the
transfer status or stops a running transfer.

Although the API itself is simple, there are several implementation details that
were very important to get right, including user authentication, parallel execution
of LCLStreamer, and concurrent management of the message buffer.
The certified package described below was used for user authentication, and the NNG-Stream
package was used as a message buffer.  A finite state machine was designed
to ensure correctness of handling all the actions involved in the transfer process.
State transitions for each transfer are driven by callbacks from
the locally running NNG-Stream and the remotely running LCLStreamer,
as well as user API calls to LCLStream-API.  Section~\ref{sec:psik}
describes the callback mechanism.

\subsection{ Message Buffering with NNG-Stream}

To support LCLStreamer's parallel data producers and consumers working,
we need a message buffer capable of aggregating
traffic from both sides at high speed.
We built \href{https://gitlab.com/frobnitzem/nng_stream}{NNG-Stream}
based on the description of the Greta/Deletria forward buffer,\cite{greta}
as their software was not available.

Placing a message buffer in-between these two provides several advantages
over direct communication: no requirement for
manually assigning producer/consumer address pairs,
resilience to delays or crashes of individual producers or consumers,
network traffic aggregation, and reduced network reachability requirements.

\begin{figure}
\includegraphics[width=2.5in]{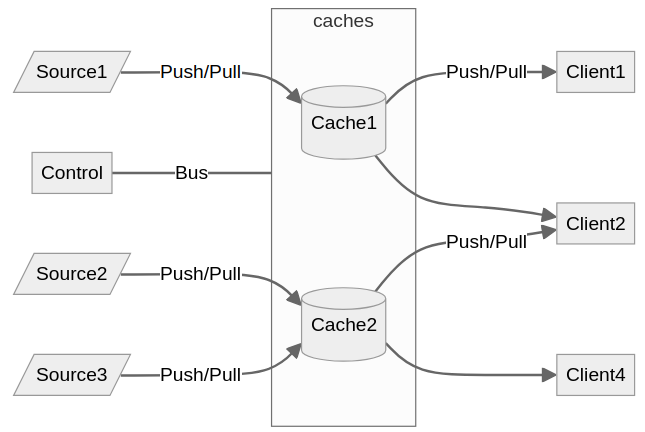}
\caption{NNG-Stream Connectivity diagram.  Each cache stores messages from all producers
in a circular buffer, and distributes them round-robin to all consumers in an at-most-once
fashion.  Connectivity is provided via NNG Push0/Pull0 socket types and, optionally ZeroMQ Push sockets.\cite{nng}
Multiple caches can work simultaneously to deliver traffic
at rates on the order of tens of gigabytes per second.
}\label{fig:nngstream}
\end{figure}

Figure~\ref{fig:nngstream} shows data flow through NNG-Stream.
Throughput tests run with a single cache on a laptop show aggregate bandwidth
of 3 Gigabytes per second. These are limited only
by local message routing and copying times.  Network link speeds are
quickly approaching the rate of 10s of Gigabytes per second.
Hence, NNG-Stream, if replicated to 3 or 4 simultaneous caches,
is capable of saturating these network links.

In our setup, NNG-Stream is run on a data transfer node
at S3DF.  These nodes serve incoming connections from
both compute nodes inside S3DF as well as compute nodes
at HPC facilities such as OLCF and NERSC.
Direct outbound traffic is technically possible for
most compute facility setups, but is sometimes blocked
administratively.  Nevertheless, this traffic path {\em{must}}
to be supported to create a useful HPC facility
working environment.  Working around administrative blocks
like this is possible by doing things like running another
NNG-Stream on a data transfer node or data streaming node
at the HPC facility.  However, setting up these extra hops
adds complexity and cost while providing
little added security benefit.

Failure properties of the data stream are inherited from
the underlying Push/Pull socket model.  Producers and consumers
can connect and disconnect from the cache without
impacting the streaming status.
Hence, producer or consumer process failures
will result in dropping the data residing in that process.
However, MPI usually terminates when a node fails,
which can cause a cascade failure.
A cache failure will terminate the stream, requiring a restart.

Normal stream shutdown is triggered by sender disconnect events.
When all senders have disconnected, the cache enters a drain
state, where no new producer connections are allowed.
When all its data has been sent, the cache disconnects
and exits.  Clients are setup to detect this disconnect
as an end-of-stream event.  Alternatively, NNG-Stream can be
configured to send ZeroMQ push data, and to send empty
frames as sentinal values on stream end.



\subsection{ Experiment Management with E-Log, ARP and Airflow}

At LCLS, the the \href{ https://pswww.slac.stanford.edu}{Electronic 
Logbook (Elog)}  is the main entry point for
users to track their experiments. It provides various services, such as
recording when and how data were collected, logging user comments and
attachments and managing a file catalog showing which files have been
written. It also includes an Automated Run Processor (ARP) that allows to
start data processing workflows without user interaction. In the Elog, users can
define processing pipelines that are launched on specific events during
the experiment (for example, when a data collection run begins or ends,
when all files with the collected data are written to a particular storage
location, etc.). The ability of the ARP to react to various different
events is made possible by the components of the LCLS data management
infrastructure exchanging information about their status via
\href{https://kafka.apache.org/}{Apache Kafka} messages.

The ARP is flexible, and can start different types of workflows tied to run events.
The two most common cases are simple shell scripts that submit jobs to a batch queue,
or sophisitcated worflows managed by an \href{https://airflow.apache.org/}{Apache Airflow} instance.
Special Airflow operators have been written to interact
with the S3DF cluster via the \href{https://slurm.schedmd.com}{SLURM} task management system, and with
the Perlmutter supercomputer at NERSC via the Superfacility API.\cite{enders_cross-facility_2020}
An additional operator that will allow communication with Psi-K (see \ref{sec:psik})
is currently under development, with the goal to allow Airflow to schedule
tasks on a higher number of HPC facilities using different task scheduling systems.

Since the E-Log/ARP integration is operational,
it is simple to enable an Airflow operator to
manage an LCLStreamer-API transfer on experiment start/stop.
This will start the data streaming pipeline as soon
as a data collection run is started.
Then compute-intensive analysis jobs will run remotely
while the experiment is taking place.

\label{sec:psik}
\subsection{ HPC Job Management with Psi-K}

The \href{https://github.com/frobnitzem/psik}{Psik} and
\href{https://github.com/frobnitzem/psik_api}{Psik-API} projects
provide a web-enabled interface for batch queuing systems
on local machines and HPC centers.
As described above, Psi-k is organized around a
file layout where a top-level directory stores a collection of
jobs. Documents within a job are rigidly structured.
Files contained within each job's folder (\code{jobs/JobID}) include:
the \code{JobSpec}, a record of job state changes, job logs,
and working directory files (as a user-managed collection).

As an example, the following Psi-k \code{JobSpec}
shows how LCLStreamer can be deployed on S3DF's
data analysis cluster.
\begin{verbatim}
JobSpec:
    name: "lclstreamer"
    directory: "/psik/76312231.123/work"
    script: 
      "mpirun -n120 lclstreamer -c cfg.yaml"
    resources:
      duration: 60 # minutes
      node_count: 1
      processes_per_node: 120
      cpu_cores_per_process: 1
    backend: S3DFslurm
    # POST to this URL on state change
    callback:
      "https://sdfdtn...edu/callbacks"
    cb_secret: "***"
\end{verbatim}
This structure combines ideas from the ExaWorks PSI/J project\cite{psij}
with traditional UNIX and cloud-based queuing systems.
Note that PSI/J is primarily a script templating and shell-based
polling engine, not designed around an asynchronous, web-API model.
Hence, it was not suitable for this work.

Jobs are queued by a POST operation, sending a JobSpec json to the
\code{jobs} path.  The server responds with either a validation error
or a new JobID.  Issuing a GET or a DELETE to \code{jobs/JobID} then
gets information about the job or cancels a queued or running job.
The result is is a single-document method for launching HPC jobs.

API security is extremely important to design from the
start.  With this in mind, all communication with the API
is strictly typed using data models, following the
\href{https://docs.pydantic.dev/latest/}{pydantic paradigm}.
In addition, more sensitive options like backend specification,
wrapper scripts, and queue options are part of
Psik-API's offline configuration.

Internally, the server names and configures each backend
(local, SLURM, API-client, etc.) using a \code{BackendConfig},
as shown below.
\begin{verbatim}
S3DFslurm: # BackendConfig
    type: slurm
    queue_name: milano
    project_name: lcls:tmox42619
\end{verbatim}
Since each \code{JobSpec} names its backend, this extra information
does not need to be a part of the API path
(differing from NERSC's Superfacility API\cite{enders_cross-facility_2020}).

This simplifies both client and server configurations.
API clients see all jobs together in a flat list, rather than separate for each backend.
On the server side, backends are logical rather than physical.
They may refer to different machines, partitions, or job scheduler attributes within a partition.

As a job, LCLStreamer is interesting because its
primary activity is to send data over a network.
This requires constant communication about the state of the
activity.  As Psi-k jobs proceed through execution phases,
they update their state files and (optionally) send callbacks.

Here is an example callback sent to inform the LCLStream-API
when the job above completes.
\begin{verbatim}
Callback:
    jobid: 76312231.123
    jobndx: 1
    state: completed
    info: ""
\end{verbatim}
Similar callbacks are sent on cancellation or failure.
These are used to manage the corresponding network data buffer
run on the data transfer node.

In addition to the above network callbacks, the job also sends
its stdout and stderr to logfiles.  These can be read from the filesystem
or, if the job is managed by Psik-API, via fetching or tailing
the logfile.

Similar to Globus Compute,\cite{globuscomput,globuscomput2} the API listener can be run by individual users
on a compute cluster's login node.  Alternatively, on Oak Ridge's Frontier system, we run it on a Kubernetes pod with network access
to the HPC cluster's job scheduler and filesystem.\cite{slate}
Oak Ridge is also implementing S3M API.\cite{s3m}
As a client application, Psi-k can send jobs to this API,
as well as to the NERSC Superfacility API\cite{enders_cross-facility_2020} and Globus compute.\cite{globuscomput2}
Differing from Globus compute, communications pass directly from the user to the Psik-API, bypasing a cloud services.

As an HTTPS server, user authentication for Psik-API can be handled several different
ways.  The simplest and most secure is to enable mutual TLS using the certified package.
This requires signing user certificates, however, and not all centers are set up
to do this at present.
It is also possible to use a reverse proxy to access Psik-API and pass
user details in the HTTP header.  OLCF Slate uses this method, since Slate provides its own HTTPS termination and authentication scheme for users based on center-issued RSA tokens.

\subsection{ Mutual authentication with Certified}

The \href{https://github.com/ORNL/certified}{Certified python package}
secures end-to-end communications with strong authentication for every
client-server and server-server interaction.  It addresses the central
challenge of secure key distribution by providing public keys.
The command-line \code{certified} program can create and sign a chain of x.509 certificates
using ed25519 public keys and signatures.  It is designed so that every python virtual environment
maintains its own separate authentication and signing key.
This way, an end-user, an application acting on the users behalf,
a microservice, and an HPC facility can all have different keys
that uniquely identify who is talking to who.

Trust in certified is established via digital signatures.
Each client is expected to obtain signatures and a list of microservices
from each organization they interact with.  The client stores those
signatures and microservice nicknames inside its configuration directory.
Then, when the client wants to issue a message to a particular HTTPS microservice,
it looks up the microservice URL and correct signature from its configuration directory.

Certified's command-line interface makes these activities simple.
Its documentation describes creating and signing certificates,
as well as managing an individual list of named, trusted microservices.
It also contains a wrapper to launch FastAPI web-servers.

Certified comes with a \code{message} command-line interface mimicking
\href{https://curl.se/}{cURL} for sending messages to REST-APIs.
Although its functionality could be duplicated with cURL,
performing the server URL translation and key lookup is complicated
and error-prone.

A companion \href{https://gitlab.com/frobnitzem/signer}{signer} package is designed
to issue signed user certificates on a UNIX/Linux system.  It can be deployed with essentially
zero configuratoin by a typical HPC facility on any login node.
Similar to \href{https://dun.github.io/munge/}{MUNGE} (the scheme for issuing
user authentication credentials for launching jobs within the \href{https://slurm.schedmd.com}{SLURM} job scheduler)
it provides assertions about the user's login name.
In detail, it takes a certificate or a certificate signing request from a user,
reads only the user's public key, and issues the user a certificate linking
their public key to their UNIX login name on that system.
The user's login name is determined by asking the kernel for
the peer's \code{SO\_PEERCRED} information.
Differing from MUNGE, it issues signed public keys.  There is no danger
of exposing these keys, since they can only be used by the original
user who posesses the corresponding private key.
Certified and signer never send the private key off of the user's device.

Both certified and signer have strong logging integration.
Certified provides a log-formatter that outputs the path accessed
and the client's identity in JSON formatting.  It can optionally be configured
to pass those logs to a \href{https://github.com/grafana/loki}{Loki server}
for viewing within \href{https://grafana.com/}{Grafana}.
The signer package stores its signatures to a database.  Signature entries
in its database can be queried for revocation status.

The combination of features above make certified a simple, viable,
drop-in authentication method for developing microservices in an
Integrated Research Infrastructure ecosystem.


\section{ Results}

This section provides details on the use of LCLStream
to address each of the application areas above.
For all of the areas, the main result to note
was that the framework successfully re-formatted LCLS data
for remote consumption and simultaneously
transmitted the data over the network to a compute
job running on OLCF.
The round-trip network latency measured between S3DF's data
transfer nodes and OLCF was consistently around 33-36 milliseconds,
only about 50\% longer than it takes light to travel the same distance.

All setups made use of parallel producers and parallel consumers.
In most cases the bottleneck (determining maximum data throughput)
was the data read/formatting speed at S3DF, around 1-3 GB/sec.
In other cases, the cache was the rate-limiting step, at around
3-4 GB/sec.  Running parallel caches could address the latter,
but was not tested in this work.

\subsection{ Masked X-ray Image Autoencoder (MAXIE) and PeakNet}

LCLS operates a variety of different detector types.
Thus, the most complicated step in this workflow is curating the
input data for training.  We produced a series of LCLStreamer
input files targeting various experiment/run IDs containing
epix10k2M images from the Psana1AssembledAreaDetector source.
We also included the photon wavelength read from psana's
\code{SIOC:SYS0:ML00:AO192} as annotation.
We opted to implement a special-purpose ``PeaknetPreprocessingPipeline"
to center and pad images to consistent sizes on the data generation
side.

In this application, the LCLStreamer input files replaced
a custom data extraction code.  The result ended up being
more readable and helped with documentation and re-use
of the curation steps.  Two primary difficulties were noted.
First, at this stage of development, it is necessary
to manually start the API on S3DF's data transfer node.
Second, we needed to implement our own client-side caching
mechanism to prevent re-downloading data.  This is significant
for this application, since ML training
makes many passes over its input.
Still, the ability to download needed inputs on-demand
decreased the data-gathering lag time between
getting an HPC allocation and being able to run scientific work.
It also eliminated stale data on the HPC side
as a potential source of errors.
These results suggest that we need robust, flexible methods
like LCLStreamer for pre-processing our data
as we work to improve its quality.

\subsection{ TMO-prefex}



The TMO-prefex application was an early version of LCLStreamer.
It was used during the commissioning of the TMO end-station
just after its upgrade to LCLS-II's higher data rate.
We used Psik-API to run a job computing electron angular
and time-of-flight correlation plots on OLCF ACE.
Then we manually started MPI-parallel TMO-prefex data producers
pushing data through NNG-Stream on S3DF.
We synchronized work by interacting through the
control room's live Zoom channel.  Due to compression at the
source (time-windowed signal thresholding), we did not
have a bandwidth limitation.
The major source of startup-latency was the program
start-up time for parallel data producers on S3DF.

\subsection{ CrystFEL}

CrystFEL has added the capability to work on live data
streamed from DECTRIS-type detectors.  The data format
is specified as part of their Simplon-API,
and consists of both control packets
and data packets.  We implemented this by adding the
\code{SimplonBinarySerializer} type as an alternative
to HDF5 serialization.  This serializer inserts the appropriate
control messages into the output stream.

During a beamtime at the Macromolecular Femtosecond Crystallography (MFX) endstation at LCLS, the LCLStreamer component was used to send data from one of the experimental runs over to the testbed HPC cluster at the Oak Ridge Leadership Computing Facility, where CrystFEL had been setup to receive and process the transferred data. The latency between data collection and processing with CrystFEL has been shown to be within the range of 15-25 seconds, allowing useful feedback information within a time range sufficient for researchers to manually steer the experiment in the desired direction.

This application especially demonstrated the benefits
obtained from standards combined with a modular,
interoperable software design.
We did not need to create a bespoke data pipeline to
generate this message format specifically,
nor did CrystFEL have to add a facility-specific data
ingestion framework.  Instead, we implemented only the
specific data format, named it after the standard, and
re-used the rest of the facility and user
community software pipelines.

\section{ Discussion}


Data streaming is an emerging technological area within
DOE's IRI (Integrated Research Infrastructure).  Our work
follows the pattern of the GRETA/Deleria project,\cite{greta}
but uses published, open-source, and modular software components
targeted to the field of X-ray image analysis.
Compared to the edge-to-exascale framework,\cite{yin24} our work
focuses on well-defined intermediate data formats and
repeatable, modular processing steps.  We also target GB/sec
data rates arising from high-frequency X-ray sources, as
opposed to somewhat lower data rates from synchrotrons
or MB/sec data rates from neutron detectors and automated
laboratories.  The latter types of data flows can still be handled
by pub/sub message transport models, as demonstrated by
INTERSECT\cite{thakur22} or BlueSky\cite{wijesinghe_bluesky_2024}
projects.
Both the latter frameworks have the somewhat larger goal
of controlling online experiments.
We do not seek to replace those frameworks, but rather to
provide a modular data request and transport mechanism.

Central to this project's success has been a continued, close collaboration between the DOE-IRI pathfinder projects, LCLS, S3DF, and leadership compute facilities.  The design boundaries between the components we eventually arrived at were shaped by sharing ideas and preliminary implementations of workflows among all these groups.
The team-developed microservices described here work in
concert to accomplish streaming.
Continued work on the IRI progam promises even greater integration
with facility-deployed infrastructure that will continue to
improve their re-usability across sites and reliability
for production work.

Metrics for measuring the pace of improvement include:
1) eliminating manual intervention steps during install, startup, and execution (such as starting servers and creating ssh tunnels) 2) reducing site-specific adaptations needed to interoperate across different experiment and compute facility API-s, 3) standardizing security technologies (such as mutual TLS authentication) needed to connect network sockets between data producers and consumers, 4) solving the "co-scheduling" problem for simultaneous commitment of beam time, network bandwidth, and compute resources, and 5) providing unencumbered access to fast, open-source data movement services (e.g. XRootD or FTS3) for managing site-to-site file copies.

\section{ Conclusion}


We have shown a new end-to-end experimental data streaming framework
incorporating key design choices for security and flexibility.
User interactions with API-s are authenticated and encrypted,
and make use of strongly typed data schemas.

It has been designed from the ground up to support new types of applications --
AI training, extremely high-rate X-ray time-of-flight analysis,
crystal structure determination with distributed processing,
and custom data science applications and visualizers yet to be created.
These are supported by a modular software stack, where individual
applications run as communicating microservices.

This project has a unique position in relation to other efforts
within the DOE Integrated Research Infrastructure (IRI) landscape.\cite{landcape24}
We have chosen to focus on small services that collaborate to provide a high-rate data channel.
This avoids dealing with the full complexity of the experimental orchestration layer that
Bluesky,\cite{wijesinghe_bluesky_2024} INTERSECT,\cite{brim_microservices_2024}
and GRETA/DELERIA's\cite{greta} Janus framework do.
At the same time, the certified, psik, and NNG-Stream packages have
given general solutions to microservice communication, HPC job submission,
and high-rate data buffering.  They are targeted to simple, drop-in
operations on HPC clusters, and avoid a centralized, cloud controller.

High-speed data streaming is a significant need for the X-ray science community.
The LCLStreamer framework has prototyped and implemented several new paradigms that meet this need.
Its successful deployment depends on the collaboration and ongoing dedication
of experts in experimental, compute, cloud service, and networking infrastructure.


\begin{acks}
We thank Daniel Pelfrey, Ross Miller, Jordan Webb, Jordan Brown, Carl Bai, and Matthew Ezell for configuring high-speed network access from the Defiant system on OLCF's ACE testbed and from compute nodes on OLCF Frontier.
This research used resources of the SLAC National Accelerator Laboratory and the Oak Ridge Leadership Computing Facility at the Oak Ridge National Laboratory, which is supported by the Office of Science of the U.S. Department of Energy under Contract No. DE-AC05-00OR22725.
Use of the Linac Coherent Light Source (LCLS), SLAC National Accelerator Laboratory, is supported by the U.S. Department of Energy, Office of Science, Office of Basic Energy Sciences under Contract No. DE-AC02-76SF00515.
\end{acks}

\bibliographystyle{plainnat}
\bibliography{bibliography}
\end{document}